\pdfoutput=1
\documentclass[reprint,amsmath,amssymb,aps,prc,twocolumn,tightenlines,superscriptaddress]{revtex4-2}
\usepackage{graphicx}
\usepackage{units}
\usepackage{bm}
\usepackage{amsfonts}
\usepackage{amsmath}
\usepackage{amssymb}
\usepackage{makecell}
\usepackage[table]{xcolor}
\usepackage{gensymb}
\usepackage[utf8]{inputenc}

\newcommand{\pic}[2][1.0]{\includegraphics[width=#1\columnwidth]{#2}}

\newcommand{\nuc}[2]{\hbox{$^{#1}$#2}}

\begin{document}


\title{Exploiting dissipative reactions to perform in-beam $\gamma$-ray spectroscopy of the neutron-deficient isotopes \nuc{38,39}{Ca}}



\author{A.\ Gade}
   \affiliation{Facility for Rare Isotope Beams,
      Michigan State University, East Lansing, Michigan 48824, USA}
   \affiliation{Department of Physics and Astronomy,
      Michigan State University, East Lansing, Michigan 48824, USA}
\author{D.\ Weisshaar}
    \affiliation{Facility for Rare Isotope Beams,
      Michigan State University, East Lansing, Michigan 48824, USA}
\author{B.\ A.\ Brown}
    \affiliation{Facility for Rare Isotope Beams,
      Michigan State University, East Lansing, Michigan 48824, USA}
    \affiliation{Department of Physics and Astronomy,
      Michigan State University, East Lansing, Michigan 48824, USA}
\author{D.\ Bazin}
    \affiliation{Facility for Rare Isotope Beams,
      Michigan State University, East Lansing, Michigan 48824, USA}
      \affiliation{Department of Physics and Astronomy,
      Michigan State University, East Lansing, Michigan 48824, USA}
\author{K.\ W.\ Brown}
    \affiliation{Facility for Rare Isotope Beams,
      Michigan State University, East Lansing, Michigan 48824, USA}
      \affiliation{Department of Chemistry,
      Michigan State University, East Lansing, Michigan 48824, USA}
\author{R.\ J.\ Charity}
    \affiliation{Department of Chemistry, Washington
    University, St. Louis, Missouri 63130, USA}
\author{P.\ Farris}
    \affiliation{Facility for Rare Isotope Beams,
      Michigan State University, East Lansing, Michigan 48824, USA}
    \affiliation{Department of Physics and Astronomy,
      Michigan State University, East Lansing, Michigan 48824, USA}
\author{A.\ M.\ Hill}
    \affiliation{Facility for Rare Isotope Beams,
      Michigan State University, East Lansing, Michigan 48824, USA}
    \affiliation{Department of Physics and Astronomy,
      Michigan State University, East Lansing, Michigan 48824, USA}
\author{J.\ Li}
    \affiliation{Facility for Rare Isotope Beams,
      Michigan State University, East Lansing, Michigan 48824, USA}
\author{B.\ Longfellow}
 \altaffiliation{Present address: Lawrence Livermore National Laboratory, Livermore, California 94550, USA}
    \affiliation{Facility for Rare Isotope Beams,
      Michigan State University, East Lansing, Michigan 48824, USA}
    \affiliation{Department of Physics and Astronomy,
      Michigan State University, East Lansing, Michigan 48824, USA}
\author{D.\ Rhodes}
\altaffiliation{Present address: TRIUMF, 4004 Wesbrook Mall, Vancouver, BC V6T 2A3, Canada}
    \affiliation{Facility for Rare Isotope Beams,
      Michigan State University, East Lansing, Michigan 48824, USA}
    \affiliation{Department of Physics and Astronomy,
      Michigan State University, East Lansing, Michigan 48824, USA}
\author{W.\ Reviol}
	\affiliation{Physics Division, Argonne National Laboratory, Argonne, Illinois 60439, USA}
\author{J.\ A.\ Tostevin}
	\affiliation{Department of Physics, Faculty of Engineering and Physical Sciences, University of Surrey, Guildford, Surrey GU2 7XH, United Kingdom}


\date{\today}

\begin{abstract}
The neutron-deficient Ca isotopes continue to attract attention due to their importance for testing isospin symmetry and their relevance in capture reactions of interest for nova nucleosynthesis and the shape of light curves in Type I X-ray bursts. To date, spectroscopic information on \nuc{38,39}{Ca} is largely limited to data on lower-spin excited states. Here, we report in-beam $\gamma$-ray spectroscopy of complementary higher-spin, complex-structure states in \nuc{39}{Ca} populated in fast-beam-induced, momentum-dissipative processes leading to neutron pickup onto excited configurations of the projectile, \nuc{9}{Be}($\nuc{38}{Ca}^*,\nuc{39}{Ca}+\gamma$)X. Such a dissipative reaction was recently characterized for the case of inelastic scattering of \nuc{38}{Ca} off \nuc{9}{Be}, \nuc{9}{Be}($\nuc{38}{Ca},
\nuc{38}{Ca}+\gamma$)X. Additional data and discussion on the nuclear structure of \nuc{38}{Ca} is also presented. An explanation for the more-complex-structure states, populated with small cross sections in one-nucleon knockout reactions, and observed in the tails of their longitudinal momentum distributions, is also offered.
\end{abstract}


\maketitle

\section{Introduction}

The study of the structure of rare isotopes is fueled by an increasing body of
complementary experimental data that reach further and further into the territory
of high isospin. Here, changes in the nuclear structure challenge nuclear theory
in the quest for a predictive model of nuclei.  Often, direct reactions or inelastic 
scattering processes are used in experimental studies that select single-particle or 
collective degrees of freedom, predominantly, providing information on low-lying states. 
Complex-configuration, higher-spin states near the yrast line that may be part of collective, band-like 
structures, have traditionally been accessed with fusion-evaporation reactions on the 
neutron-deficient side of the nuclear chart when suitable stable projectile and target 
combinations exist. Incomplete fusion~\cite{dra97} or cluster transfer reactions in normal kinematics induced by stable beams \cite{hae82} or inverse-kinematics reactions with radioactive projectiles and light stable targets \cite{ide05,bot15,ash21}, at low beam energies, have 
been used in only a very few cases to access high-spin yrast states; the latter type of studies is sparse
due to the limited availability of the intense low-energy rare-isotope beams that 
are necessary.

\begin{figure}[h]
\begin{center}
\pic{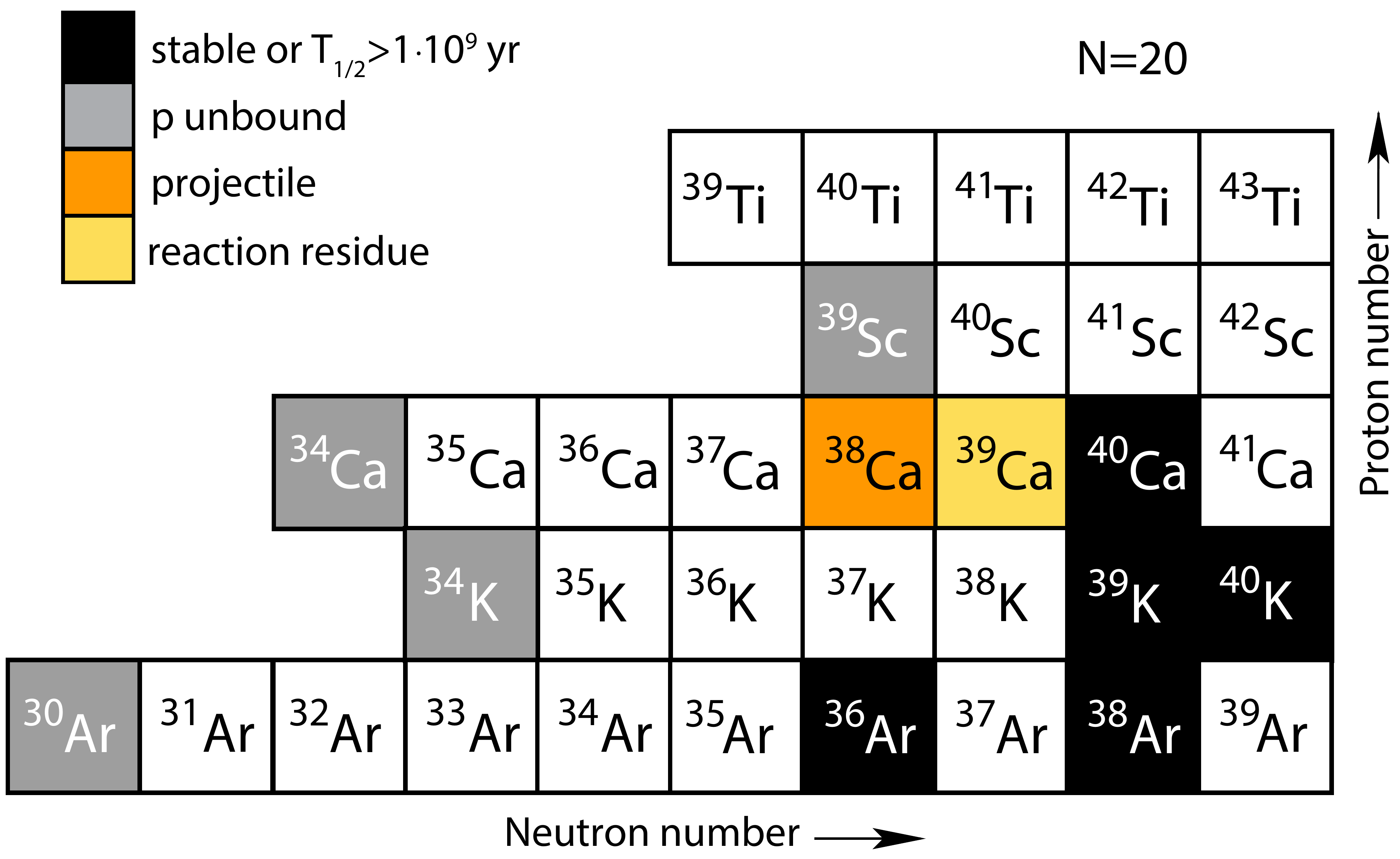}
\caption{Part of the nuclear chart showing the projectile beam \nuc{38}{Ca} and the one-neutron
pickup residue \nuc{39}{Ca} in the proximity of the proton dripline. }
\label{fig:chart}
\end{center}
\end{figure}

Here, we extend the work on neutron-deficient nuclei reported in Ref. \cite{gad22} and report the in-beam $\gamma$-ray 
spectroscopy of complex-structure, core-coupled states near the yrast line of \nuc{39}{Ca}. These states were populated in 
fast-beam-induced, dissipative (high-momentum-loss) reaction processes leading to 
single-neutron addition onto excited configurations of the projectile, \nuc{9}{Be}($
\nuc{38}{Ca}^*,\nuc{39}{Ca}+\gamma$)X. Complementing the analysis presented in Ref. 
\cite{gad22}, we provide further discussion on the inelastic scattering, \nuc{9}{Be}($
\nuc{38}{Ca},\nuc{38}{Ca}+\gamma$)X, at high momentum loss. Figure~\ref{fig:chart} shows 
the location of these systems on the nuclear chart.

The neutron-deficient \nuc{38}{Ca} isotope continues to attract attention due to its
importance for testing fundamental symmetries \cite{bol06,rin07,geo07,par11,par14,par15,bla15}
and as the compound nucleus in the \nuc{34}{Ar}$(\alpha,p)$\nuc{37}{K} capture reaction of
relevance for X-ray bursts~\cite{lon17}. For nuclear structure studies, \nuc{38}{Ca}
can be accessed via (\nuc{3}{He},$n$) and $(p,t)$ reactions from stable \nuc{36}{Ar} and
\nuc{40}{Ca}, respectively. Hence an extensive body of work is available on, for example, pairing
vibrations~\cite{boh77}, $fp$-shell configurations \cite{alf86}, and an anomalous $L=0$
transition to the first excited $0^+$ state. The latter was observed in the $(p,t)$ reaction
from \nuc{40}{Ca}~\cite{har66, kub77,kub74, pad72}. The intermediate-energy Coulomb excitation
measurement by Cottle {\it et al.}~\cite{cot99}, exploring isospin symmetry, is the only one
using a beam of unstable \nuc{38}{Ca} projectiles. The sole published studies using $\gamma$-ray
spectroscopy are from 1970 \cite{sha70} ($(\nuc{3}{He},n\gamma)$, using three Ge(Li) detectors)
and 1999 \cite{cot99} (intermediate-energy Coulomb excitation, using a NaI scintillator array).
NNDC further quotes $\gamma$-ray data from a 1974 Duke University PhD Thesis \cite{nndc38} which
appears unpublished for all practical purposes. The data presented here constitute the first
$\gamma$-ray spectroscopy of \nuc{38}{Ca} exploiting a modern, high-resolution HPGe $\gamma$-ray
tracking array.

Unlike for \nuc{38}{Ca}, modern $\gamma$-ray spectroscopy data are available for \nuc{39}{Ca}.
For example, from (a) a recent $(\nuc{3}{He},\alpha\gamma)$ experiment~\cite{hal20}, aimed at
observables important for the \nuc{38}{K}$(p,\gamma)$\nuc{39}{Ca} reaction rate, and (b) from a
high-spin spectroscopy study via the \nuc{16}{O}$(\nuc{28}{Si},\alpha n \gamma)$\nuc{39}{Ca}
reaction~\cite{and99}. In the present work, we show that dissipative processes in the fast-beam
one-neutron pickup channel populate the very same states as reported in the high-spin study
of Ref. \cite{and99}, suggesting a possible novel and practical pathway to study such states
in rare isotopes. This also elucidates the observation of such states, with low yields, in the 
low-momentum tails of longitudinal momentum distributions in nucleon knockout experiments, 
reported in recent measurements.

 \section{Experiment, Results and Discussion}

\begin{figure}[h]
  \begin{center}
    \pic{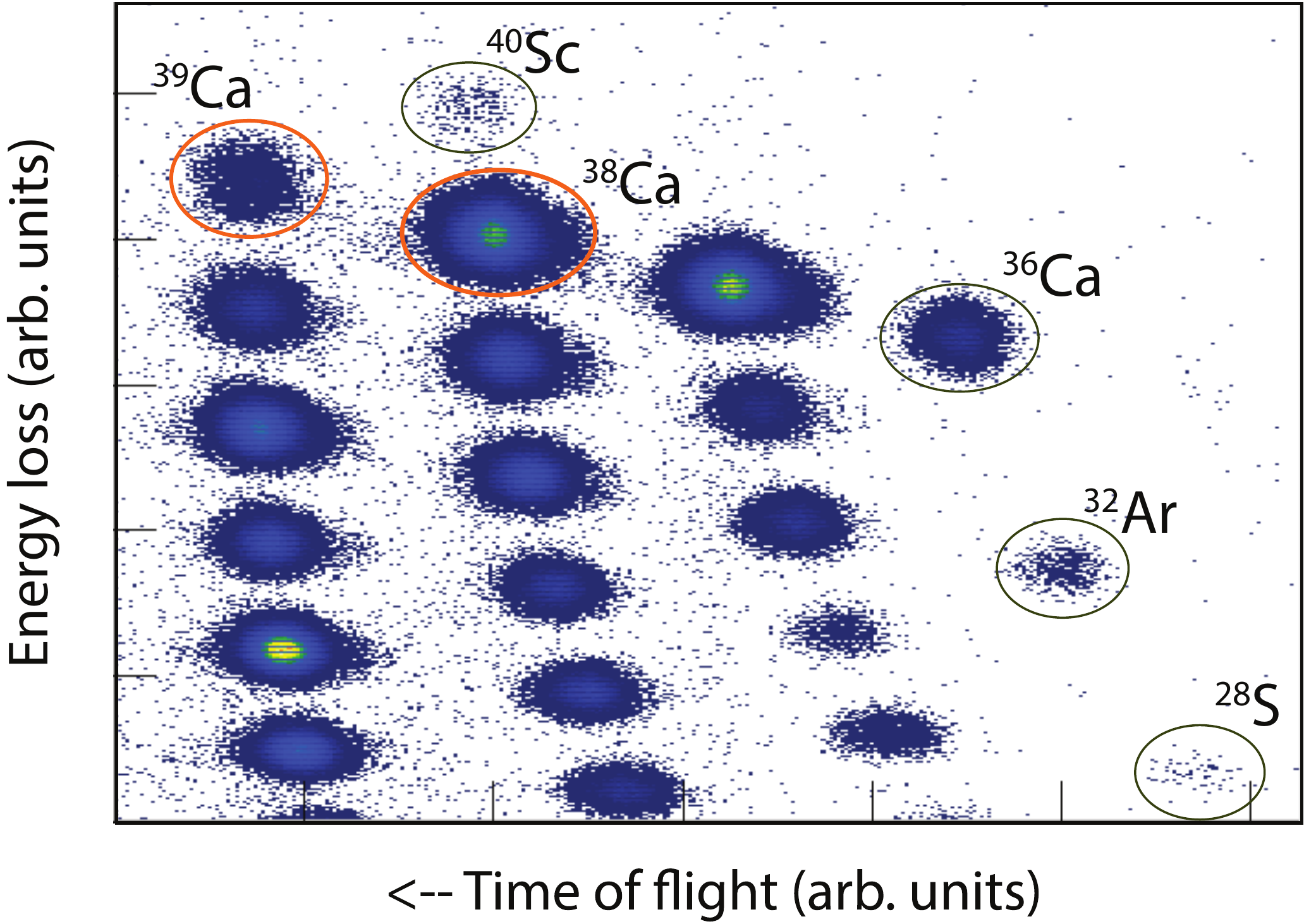}
    \caption{Particle identification showing the same data as in Fig. 2 of Ref. \cite{gad20} with \nuc{38,39}{Ca} highlighted.}
    \label{fig:pid}
  \end{center}
\end{figure}

The experimental details and setup are also discussed in Refs. \cite{gad20,gad22} and a brief summary is provided here. The \nuc{38}{Ca} rare-isotope beam was produced at the Coupled Cyclotron Facility at NSCL \cite{nscl} in the fragmentation of a stable 140-MeV/nucleon \nuc{40}{Ca} primary beam in the A1900 fragment separator \cite{a1900}, on a 799 mg/cm$^2$ \nuc{9}{Be} production target and separated using a 300 mg/cm$^2$ Al degrader. The momentum width was limited to $\Delta p/p =0.25$\% for optimum resolution, resulting in 160,000 \nuc{38}{Ca}/s interacting with a 188-mg/cm$^2$-thick \nuc{9}{Be} foil placed in the center of the high-resolution $\gamma$-ray tracking array GRETINA~\cite{pas13,wei17} surrounding the reaction target position of the S800 spectrograph \cite{s800}. The \nuc{38}{Ca} projectiles had a mid-target energy of 60.9~MeV/nucleon. The incoming beam and the projectile-like reaction residues were event-by-event identified using the S800 analysis beam line and focal plane \cite{s800FP}. The scattered \nuc{38}{Ca} and one-neutron pickup residues, \nuc{39}{Ca}, are cleanly separated in the particle identification plot displayed in Fig. \ref{fig:pid}. The magnetic rigidity of the S800 spectrograph was tuned for the two-neutron removal reaction to \nuc{36}{Ca} and so only the outermost low-momentum tails of the reacted \nuc{38}{Ca} and \nuc{39}{Ca} longitudinal momentum distributions were transmitted to the S800 focal plane, as quantified below.  In the following, we  will first discuss the general characteristics of the reactions and then focus on the detailed spectroscopy results for \nuc{39}{Ca} and on additional data and discussion for \nuc{38}{Ca} not presented in \cite{gad22}. 

\begin{figure}[htb]
\begin{center}
\pic{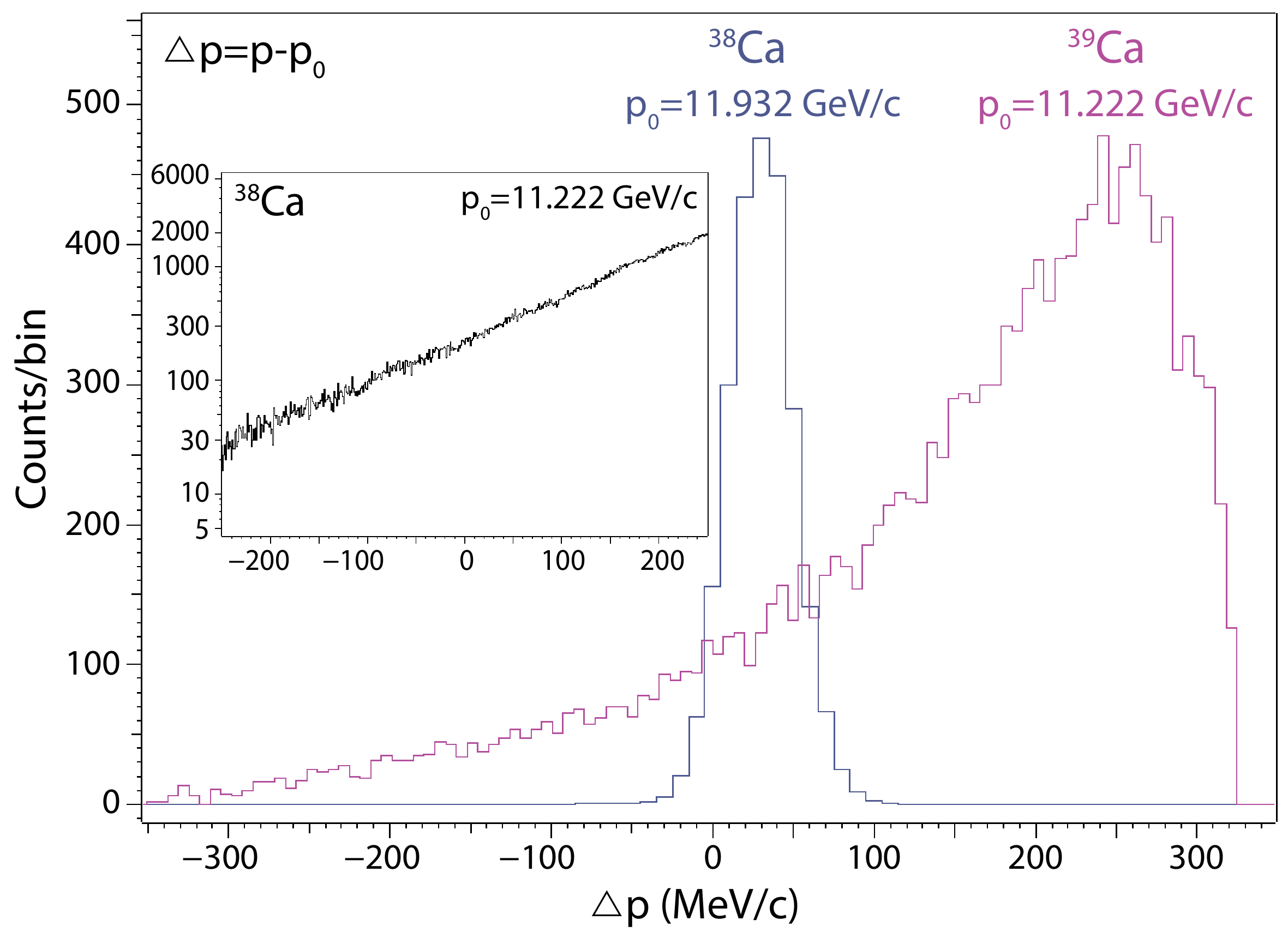}
\caption{Measured tails of the parallel momentum distributions, versus $\Delta p =p -
p_0$ (with $p_0=11.222$ GeV/c) for the \nuc{39}{Ca} one-neutron pickup residues (purple
curve) and, as an inset, for the \nuc{38}{Ca} nuclei that have undergone a momentum
loss of order of 700~MeV/c (black curve). The latter is shown on a logarithmic scale,
revealing its exponential character. The momentum distribution versus $\Delta p =p - p_0$
(with $p_0=11.932$ GeV/c) of the unreacted \nuc{38}{Ca} after passage through the
target is also shown (blue curve). These central momentum values, $p_0$, for each isotope,
correspond to the respective magnetic rigidity setting of the S800 spectrograph.
The low-momentum tail distributions of the \nuc{39}{Ca} reaction residues and of the
reacted \nuc{38}{Ca} entered the focal plane in the same setting and thus $p_0$ value, while the unreacted \nuc{38}{Ca} is from a different magnetic setting with a different $p_0$ value.}
\label{fig:39Ca_ppar}
\end{center}
\end{figure}

With the chosen magnetic rigidity, optimized for \nuc{36}{Ca}, the parts
of the \nuc{39}{Ca} and \nuc{38}{Ca} momentum distributions that enter the S800 focal
plane are approximately $\pm 300$ MeV/c about the momentum $p_0=11.222$ GeV/c. Figure
\ref{fig:39Ca_ppar} shows the measured parallel momentum distributions of (i) the
low-momentum tail of the \nuc{38}{Ca} distribution on a logarithmic scale (shown as
an inset), over a slightly reduced momentum range that is not subject to further
acceptance effects, and (ii) the tail of the \nuc{39}{Ca} one-neutron pickup distribution
(purple curve) from the same setting. This \nuc{39}{Ca} distribution is further impacted
by the focal-plane acceptance starting at about $p_0+275$~MeV/c. The momentum distribution
of the unreacted \nuc{38}{Ca} after passage through the target is also shown (blue curve,
roughly centered on $p_0=11.932$ GeV/c) to help clarify the momentum losses in the 
observed \nuc{39}{Ca} and \nuc{38}{Ca} tail distributions.

The parallel momentum distribution of \nuc{39}{Ca} is cut by the S800 focal-plane
acceptance on the high-momentum end, leaving only a tail within the acceptance. It
is interesting to estimate where the centroid of the full distribution would be. The
complication here is that this depends on the momentum of the neutron picked up from
the target which is not precisely calculable, given that our reaction is highly
linear-momentum mismatched and is dominated by the pickup of deeply bound, high-momentum
neutrons \cite{gad11}. Such fast-beam one-nucleon pickup reactions have been carried
out for a number of projectiles upon \nuc{12}{C} and \nuc{9}{Be} targets within the
same experimental scheme~\cite{gad11,gad07,gad08,mcd12,gad16}.  The only one-neutron
pickup from a \nuc{9}{Be} target onto a neutron-deficient projectile was $\nuc{9}{Be}
(\nuc{22}{Mg}, \nuc{23}{Mg}+\gamma)X$.

So, to estimate the centroid position of the full distribution we resort to the
systematics presented in Fig. 7 of Soulioutis {\it et al.} \cite{sou92}.
This confronts the centroid of the parallel momentum per nucleon of the pickup
residues relative to the parallel momentum per nucleon of the beam, $f$, with the
number of nucleons transferred from the target, $\Delta A_t$, relative to the mass
of the fragment, $A_f$: i.e. $f$ vs. $\Delta A_t/A_f$. As one would expect, the
systematics show that with increasing fragment mass and a small number of nucleons
transferred, $f$ approaches 1, meaning that the parallel momentum per nucleon between
the beam and pickup residues is essentially unchanged following the reaction. However,
the systematics include only pickup from heavier targets, such as Al and Ta \cite{sou92},
and it is not {\em a priori} clear how well these may apply to the neutron pickup from
the complex Be target -- which may be viewed as having predominantly the structure of
two $\alpha$ particles plus an additional neutron. To benchmark the systematics for a
one-neutron pickup from \nuc{9}{Be} onto a neutron-deficient projectile, we deduce
$f=0.974$ from Fig.~3 of \cite{gad11} and extrapolation to a mid-target reaction
vertex for $\nuc{9}{Be}(\nuc{22}{Mg}, \nuc{23}{Mg}+\gamma)X$. This result indeed
fits well into the systematics of \cite{sou92} at $\Delta A_t/A_f=0.04$. Now,
for \nuc{39}{Ca} and $\Delta A_t/A_f=0.026$, an $f$ closer to 0.990 would be
expected. Using this value to calculate the parallel momentum of \nuc{39}{Ca} from
pickup onto \nuc{38}{Ca} at mid-target predicts a centroid of the parallel momentum
distribution, as measured behind the target, of 12.165~GeV/c, more than 600~MeV/c
outside of the S800 acceptance captured by our setting. So, indeed, we appear to
observe only the outermost tail of the \nuc{39}{Ca} pickup distribution. Although
the fast-beam pickup reactions are very sensitive to the nuclear structure of the
projectile and the beam energy, perhaps supporting that only the outermost tail is
observed, the cross section determined for the part of the distribution we observe
--  from the number of \nuc{39}{Ca} relative to the number of \nuc{38}{Ca} projectiles
and target nuclei -- yields $\sigma(p_0 \pm 330 \mathrm{MeV/c})=0.11(1)$~mb, which
is more than a factor of 20 smaller than the inclusive cross section for the
$\nuc{22}{Mg}+n$ reaction reported in Ref. \cite{gad11}, for example.

The low-momentum tails of parallel momentum distributions in fast-beam, one-nucleon 
pickup, as well as in knockout reactions, have long been suspected to reveal the presence 
of rare, higher-order, more-dissipative processes \cite{gad08,str14,mut16,cra17,spi19}.
Of particular relevance may be the observation of a core-coupled state in the $\nuc{9}{Be}
(\nuc{22}{Mg}, \nuc{23}{Al}+\gamma)X$ one-proton pickup reaction where a pronounced
low-momentum tail dominated the measured parallel momentum distribution of \nuc{23}{Al}
~\cite{gad08}. There, the only excited state observed was the first $(7/2^+)$ level,
almost 1400 keV above the proton separation energy, which could only be reached in
a complex rearrangement: pickup of a $d_{5/2}$ proton onto \nuc{22}{Mg} excited into the $2^+_1$
state \cite{gad08}. This was indeed supported by the fact that the excitation of the
$2^+_1$ and $4^+_1$ states was observed in the low-momentum tail of the reacted \nuc{22}{Mg}
projectiles passing through the target~\cite{gad08}. Our current observations display an
astonishing bounty of these more-complex, higher-spin states, accessible in \nuc{39}{Ca}
via in-beam $\gamma$-ray spectroscopy. The spectroscopy of \nuc{39}{Ca} will be discussed
in the following subsection.

When compared to the parallel momentum distribution of the unreacted \nuc{38}{Ca}
passing through the target and suffering only in-target energy loss, the reacted
\nuc{38}{Ca} nuclei in the reaction setting have undergone an additional longitudinal
momentum loss of about 700~MeV/c, that is 18~MeV/c per nucleon in momentum or
5.4~MeV/nucleon in energy. The logarithmic plot in the inset in Fig. \ref{fig:39Ca_ppar} 
reveals that this low-momentum tail of the \nuc{38}{Ca} parallel momentum distribution
is essentially exponential. As for the pickup cross section quoted above, the cross section
for finding \nuc{38}{Ca} with large momentum loss in the interval $p_0 \pm 330$~MeV/c was
extracted to yield $\sigma(p_0 \pm 330~\mathrm{MeV/c})=3.8(4)$~mb, making these inelastic
large-momentum-loss events also rather rare. See Ref. \cite{gad22} for a detailed discussion 
of the most strongly populated configurations in \nuc{38}{Ca} at high momentum loss.

\subsection{\nuc{39}{Ca} spectroscopy}

\begin{figure}[h]
\begin{center}
\pic{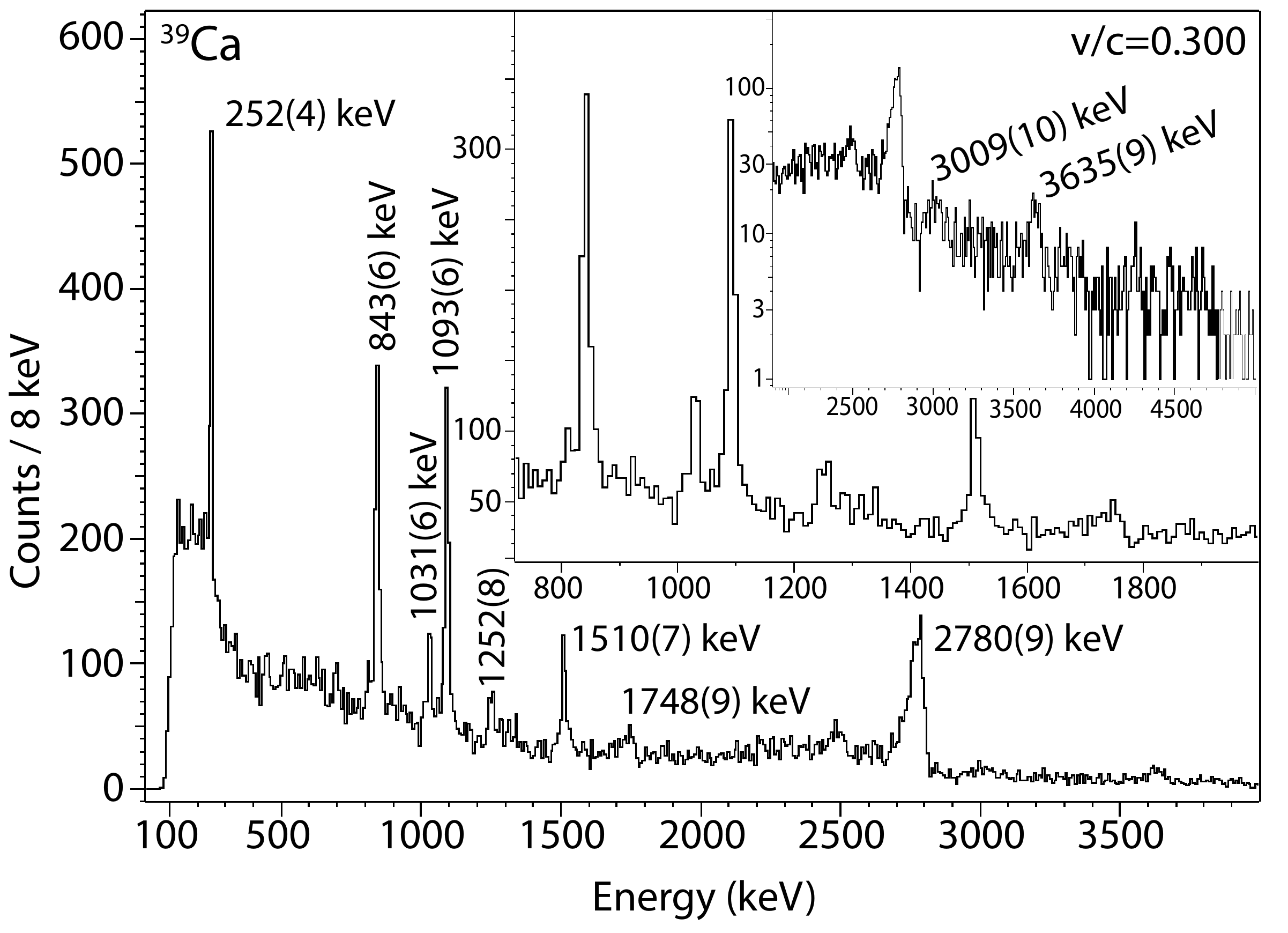}
\caption{Doppler-reconstructed addback $\gamma$-ray spectrum as detected in coincidence
with the \nuc{39}{Ca} one-neutron pickup residues identified in the S800 focal plane.
All $\gamma$-ray transitions are labeled by their energy. The two insets magnify
low-statistics regions of the spectrum. All transitions are known -- nine of the ten
reported energies agree with the literature values within uncertainties and the
discrepancy for the 2797-keV transition \cite{and99} is attributed to lifetime
effects as discussed in the text.}
\label{fig:Ca39_sing}
\end{center}
\end{figure}

The event-by-event Doppler reconstructed $\gamma$-ray spectrum measured in coincidence
with the \nuc{39}{Ca} reaction residues detected in the S800 focal plane is shown in
Fig.~\ref{fig:Ca39_sing}. Nearest neighbor addback, as detailed in \cite{wei17}, was used.
All $\gamma$-ray transitions observed here have been reported before, for example in
Refs. \cite{and99} and \cite{hal20}, with unambiguous placements in the \nuc{39}{Ca}
level scheme. A few peculiarities of in-beam $\gamma$-ray spectroscopy at or above
30\% of the speed of light are apparent. The $v/c=0.30$ chosen for the Doppler
reconstruction gives the best resolution for the presumably fast $M1/E2$ transitions,
such as at 252, 843, 1031 and 1093~keV, while the peak quoted at 2780~keV displays a
pronounced left-side tail and is too low in energy as compared to the literature
value of 2796~keV \cite{hal20}. This is due to the corresponding $7/2^-$ excited-state
having a long half-life of 62(17)~ps \cite{nndc39}, thus decaying well behind the target
and being Doppler-reconstructed to an energy value that is too low. This effect was
both discussed and simulated for this beam energy regime and setup in Ref. \cite{gad19}
for the case of \nuc{70}{Fe}.

It is noted that there is no clear evidence for the 2467-keV $1/2^+$ first excited
state, which would also be obscured by the Compton edge of the 2780-keV $\gamma$ ray
visible at about 2500~keV. Indeed, the population of this state would not be expected
in fast-beam one-neutron pickup due to the large momentum mismatch at our high beam
energies. Table~\ref{tab:intensity39} provides the observed relative $\gamma$-ray
intensities in \nuc{39}{Ca}. These are obtained from the efficiency-corrected peak
areas and include a 7\% systematic error added in quadrature to account for uncertainties
in the efficiency when including addback.

\begin{table}[ht]
\caption{Relative intensities $I_{\gamma}$ of the $\gamma$-ray transitions with energy
$E_{\gamma}$ extracted from the \nuc{39}{Ca} $\gamma$-ray spectrum shown in Fig.~\ref{fig:Ca39_sing}.}
\begin{ruledtabular}
\begin{tabular}{rr}
$E_{\gamma}$ (keV) & $I_{\gamma}$ (\%) \\
\hline
  252(4) & 20(3)\\
  843(6) &  41(4)\\
  1031(6) & 15(2)\\
  1093(6) &  41(4)\\
  1252(8) & 9(3)\\
  1510(7) &  18(3)\\
  1748(9) &  3(1)\\
  2780(9) & 100(10)\\
  3009(10) &  12(3)\\
  3635(9)&  15(3)\\
\end{tabular}
\end{ruledtabular}
\label{tab:intensity39}
\end{table}

\begin{figure}[h]
\begin{center}
\pic[0.60]{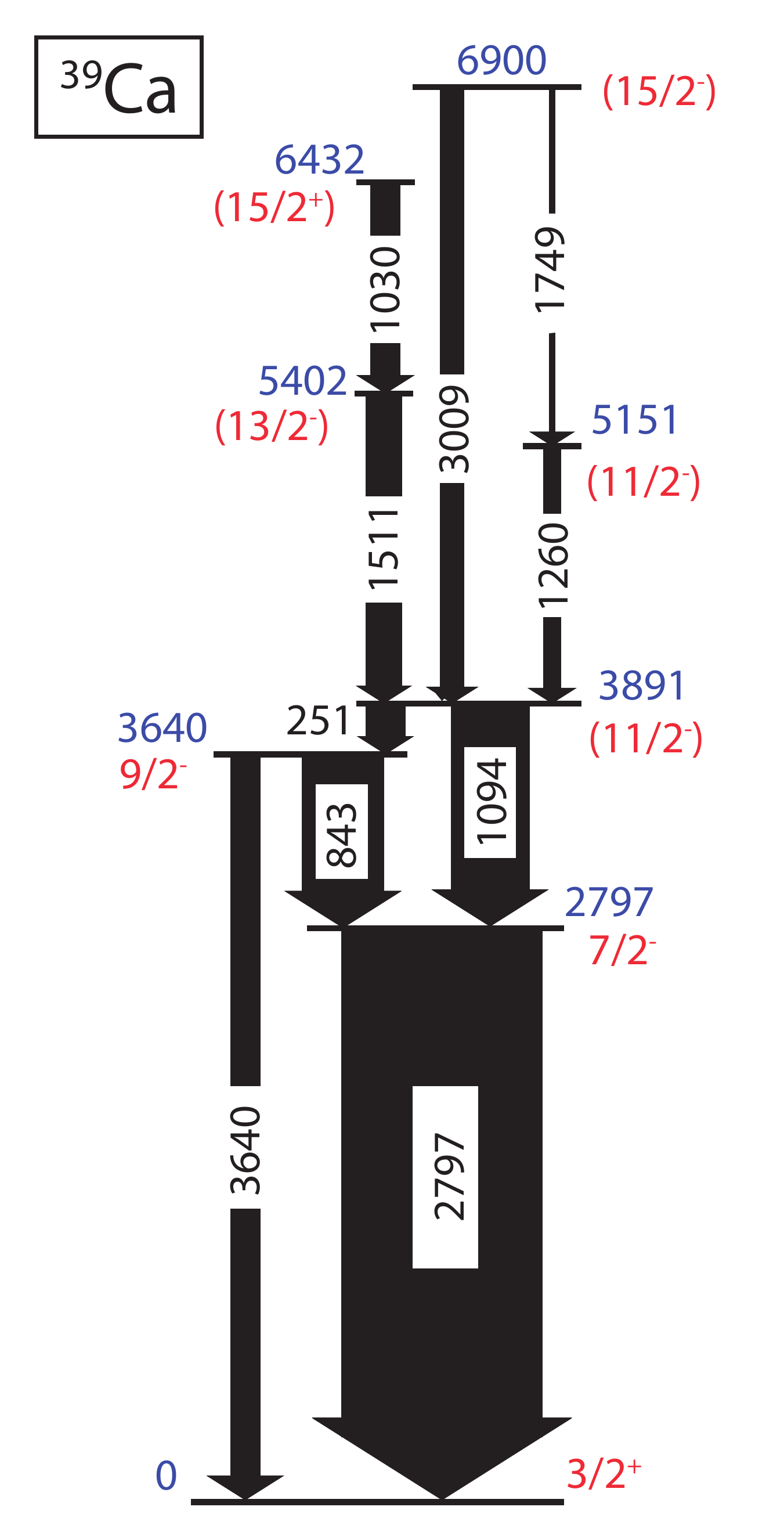}
\caption{Level scheme of \nuc{39}{Ca} as populated in this work. The line width of the
arrows corresponds to the intensities of the transitions quoted in Table \ref{tab:intensity39}.
We note that the \nuc{39}{Ca} proton separation energy is $S_p=5.7709(6)$~MeV~\cite{ame20},
placing the $J=15/2$ states above this value. For the purpose of this level scheme, we use
the energies quoted in Ref. \cite{and99}, which, except for the 2797-keV transition due to lifetime effects as discussed in the text, agree within uncertainties with the present work. } 
\label{fig:Ca39_lev}
\end{center}
\end{figure}

Figure \ref{fig:Ca39_lev} displays the level scheme of \nuc{39}{Ca} reported from the
present work where the width of the arrows scale with the $\gamma$-ray intensities
quoted in Table \ref{tab:intensity39}. It is noteworthy that the level scheme populated
here is essentially identical to that reported in Ref. \cite{and99} from the \nuc{28}{Si}
 + \nuc{16}{O} fusion-evaporation reaction. The only state and corresponding transition
reported in Fig. 3(b) of \cite{and99} that is not observed here is a proposed ($19/2^-$)
state above 7.7~MeV in excitation energy.

\begin{figure}[h]
  \begin{center}
    \pic{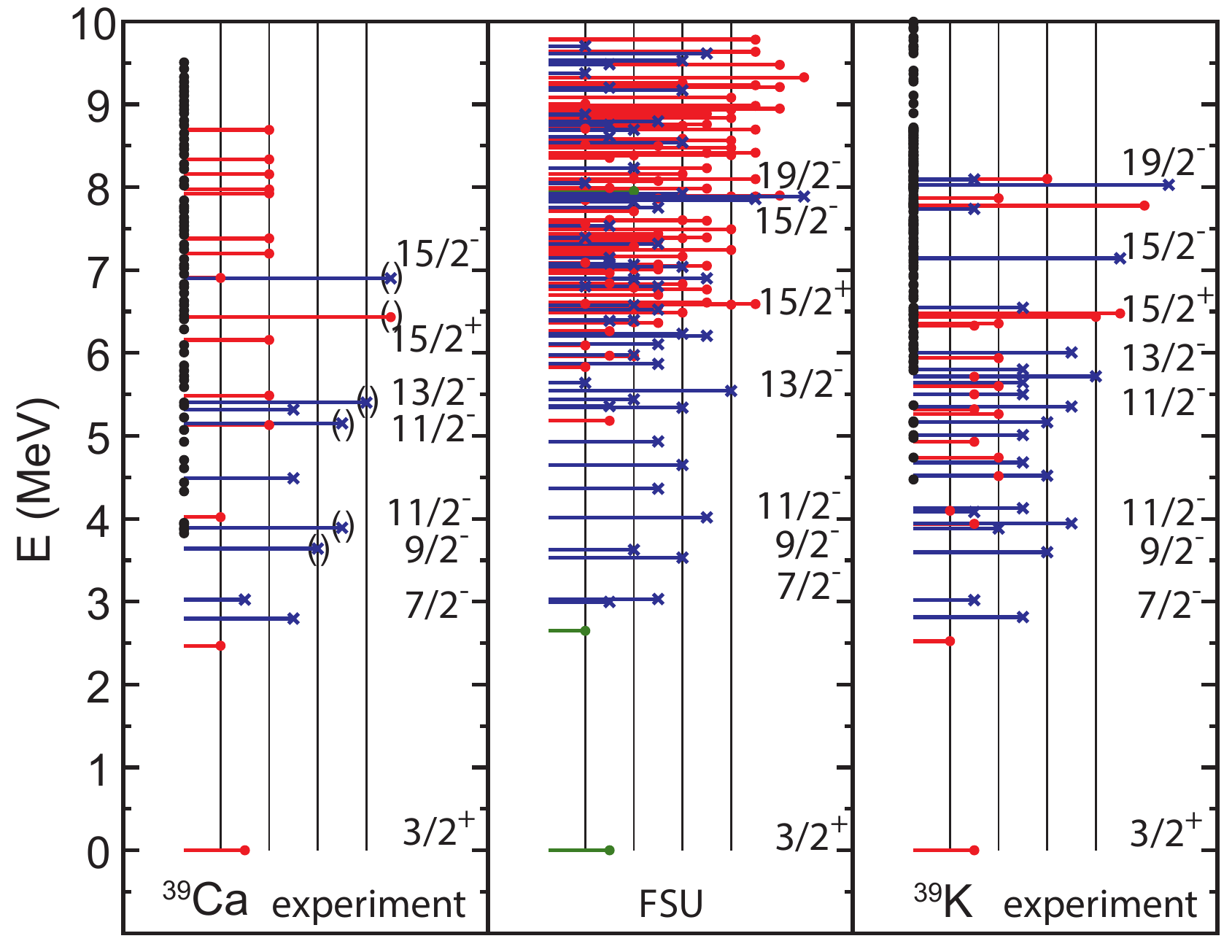}
    \caption{Comparison of the energies of the low-lying states of \nuc{39}{Ca}, with the states observed here labeled, with shell-model calculations using the FSU $s psd fp$ interaction and the mirror nucleus \nuc{39}{K} \cite{nndc39}. In these plots, the length of the levels indicates the $J$ value and the color positive parity (red), negative parity (blue), and $sd$-shell origin (green). For the panels showing experimental data, the black dots indicate levels with unknown spin-parity assignment.  }
    \label{fig:Ca39_theo}
  \end{center}
\end{figure}

These observations are in contrast to the population of final states in the \nuc{40}{Ca}(
\nuc{3}{He},$\alpha$)\nuc{39}{Ca} transfer reactions of Refs. \cite{hal20,set18}. There, a
number of strong positive- and negative-parity low-spin states were reported in addition to
some of the high-spin, more-complex-structure states seen here but which appear only very
weakly populated in Fig. 2 of Ref. \cite{set18}, for example.

From Ref. \cite{and99}, relative to the $0p-0h$ configurations of the \nuc{40}{Ca} ground 
state, the configurations of the states observed in the \nuc{39}{Ca} level
scheme are dominated by (i) $0p-1h$ for the $3/2^+$ ground state, (ii) $1p-2h$ for the $7/2^-$
to $13/2^-$ negative-parity multiplet, (iii)  $2p-3h$ for the $15/2^+$ state (one of the lowest-lying
members of the stretched $\pi(f_{7/2})\otimes \nu(f_{7/2})$ $7^+$ configuration coupled to
the $K=1/2$ component of the $d_{3/2}^{-3}$ configuration), and (iv) $3p-4h$ for the $15/2^-$
state which is calculated to be one of the lowest-lying members of the resulting multiplet. The present calculations for $3p-4h$ are limited to the $15/2^-$ and $19/2^-$ states shown in Fig.~\ref{fig:Ca39_theo}.

One might be tempted to consider complex two-step production mechanisms on different target nuclei, such as proton removal
from \nuc{38}{Ca} with subsequent $pn$ pickup, but estimates are that these would have a combined
cross section orders of magnitude too small -- the $pn$ pickup cross section was determined to be
of order $\mu b$, as reported in \cite{gad20}. In the absence of more exclusive data to elucidate
the reaction mechanisms, it appears reasonable to speculate that the population of the observed
states involves one-neutron pickup onto a \nuc{38}{Ca} core that has undergone single or multiple
nucleon rearrangements relative to the ground-state configuration.

While the above-mentioned dominant shell-model configurations originate from the calculations of 
Ref. \cite{and99} -- in the very truncated $d_{3/2}-f_{7/2}$ model space but allowing for all 
configurations of seven valence particles -- it is interesting to compare the data also to new 
calculations in an extended model space ($s psd fp$) using a modern cross-shell interaction 
(FSU) \cite{lub19,lub20} restricted to pure $\hbar \omega$ configurations. The comparison is 
shown in Fig. \ref{fig:Ca39_theo} and shows a remarkable agreement between the measured \nuc{39}{Ca} 
energies and the calculations and highlights the mirror symmetry between Ca and K at $A=39$.  
In the $A=39$ mirror system, the second $11/2^-$ state near 5~MeV and
first $15/2^-$ state near 7~MeV are about 1~MeV lower in energy than that calculated.
The close spacing of the $15/2^-$ and $19/2^-$ in the calculation is a reflection of the close spacing
predicted and observed for the ($fp$)$^3$ configuration in \nuc{43}{Sc} and \nuc{43}{Ti}.
The more rotational like spacing of the experimental $11/2_2^-$, $15/2^-$ and $19/2^-$ energies in \nuc{39}{K} may be due to
deformation beyond that contained in the $3p-4h$ truncation.

\subsection{\nuc{38}{Ca} spectroscopy}

Complementing the spectroscopy data and shell-model theory discussion published in Ref. \cite{gad22}, presented here is the full level scheme including very weak transitions and the corresponding coincidence relationships not discussed in \cite{gad22}, all relative peak intensities with their uncertainties, and the determination of an upper limit for the $3^-_1 \rightarrow 0^+_1$ branch which is relevant for Coulomb excitation studies that report population of the near-degenerate $3^-_1$ and $2^+_2$ states.

\begin{figure}[h]
\begin{center}
\pic[0.95]{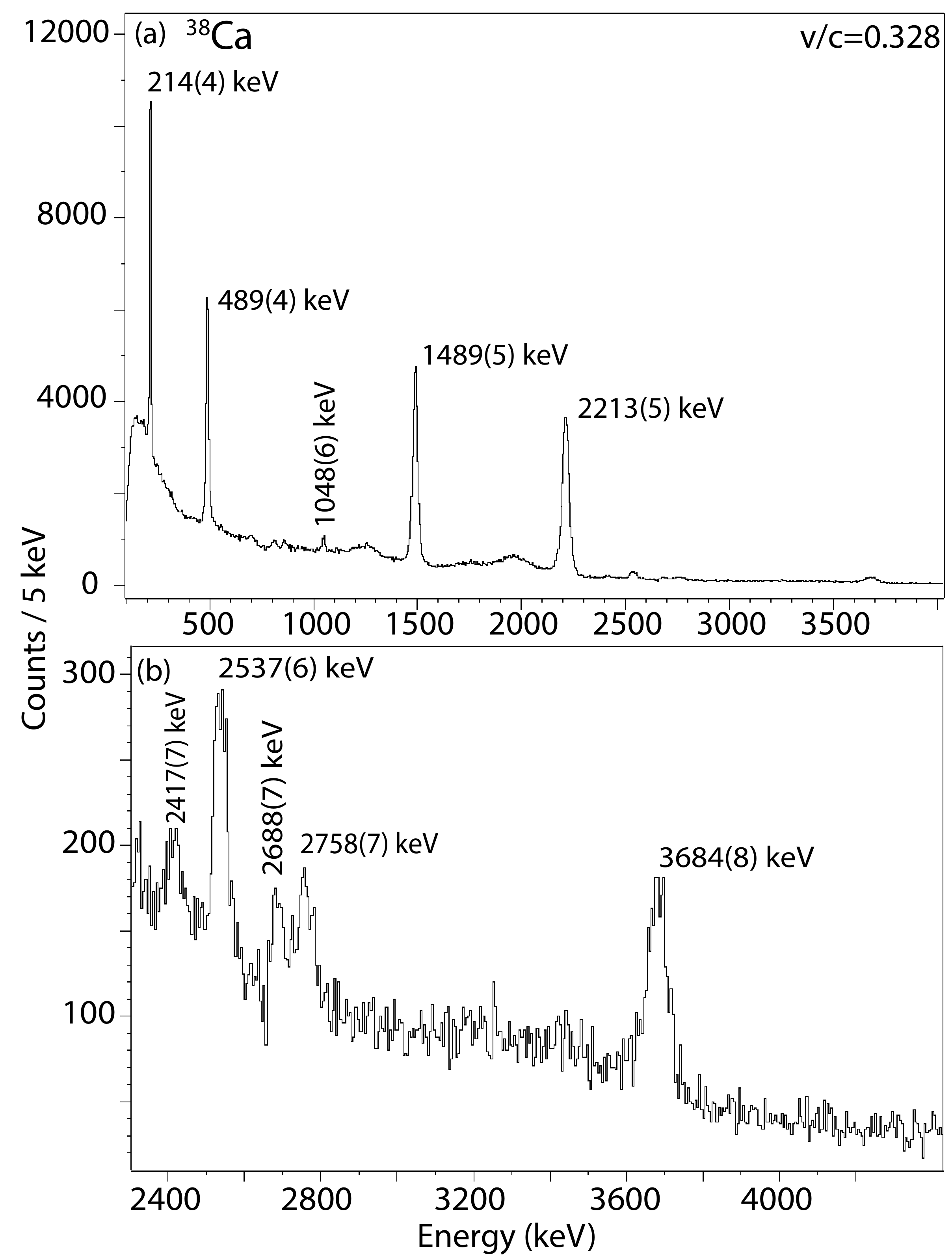}
\caption{Doppler-reconstructed addback $\gamma$-ray spectrum as detected in coincidence with
the scattered \nuc{38}{Ca} nuclei that underwent a large momentum loss. All $\gamma$-ray
transitions are labeled by their energy. Panels (a) and (b) show the low-energy and high-energy region of the
spectrum, respectively. }
\label{fig:Ca38_sing}
\end{center}
\end{figure}

The event-by-event Doppler reconstructed $\gamma$-ray spectrum obtained in coincidence with
the \nuc{38}{Ca} reaction residues detected in the S800 focal plane at large momentum loss is
shown in Fig.~\ref{fig:Ca38_sing}. Nearest neighbor addback, as detailed in \cite{wei17}, was
used. Of the seven $\gamma$-ray transitions compiled in \cite{nndc38}, those at 2213(5), 1489(5),
489(4), 3684(8)~keV are observed here, while the transitions at 214(4), 1048(6), 2417(7), 2537(6),
2688(7), and 2758(7)~keV are reported for the first time in the present work (see also \cite{gad22}).

Table~\ref{tab:intensity38} lists the relative $\gamma$-ray intensities deduced from the
efficiency-corrected peak areas from the spectrum displayed in Fig.~\ref{fig:Ca38_sing}.
These include a 7\% error, added in quadrature, to account for the uncertainty in the
photopeak efficiency with addback. Remarkably, the fourth strongest $\gamma$ ray, at
214~keV, has not been reported previously.

\begin{table}[ht]
\caption{Relative intensities $I_{\gamma}$ of the $\gamma$-ray transitions with energy
$E_{\gamma}$ extracted from the \nuc{38}{Ca} $\gamma$-ray spectrum shown in Fig.~\ref{fig:Ca38_sing}.}
\begin{ruledtabular}
\begin{tabular}{rr}
$E_{\gamma}$ (keV) & $I_{\gamma}$ (\%) \\
\hline
  214(4) &  27(2)\\
  489(4) &  31(2)\\
  1048(6) & 3.0(5)\\
  1489(5) &  75(5)\\
  2213(5) & 100(7)\\
  2417(7) &  1.8(3)\\
  2537(6) &  4.8(5)\\
  2688(7) & 2.0(3)\\
  2758(7) &  2.9(4)\\
  3684(8) &  6.6(6)\\
\end{tabular}
\end{ruledtabular}
\label{tab:intensity38}
\end{table}

\begin{figure}[h]
\begin{center}
\pic{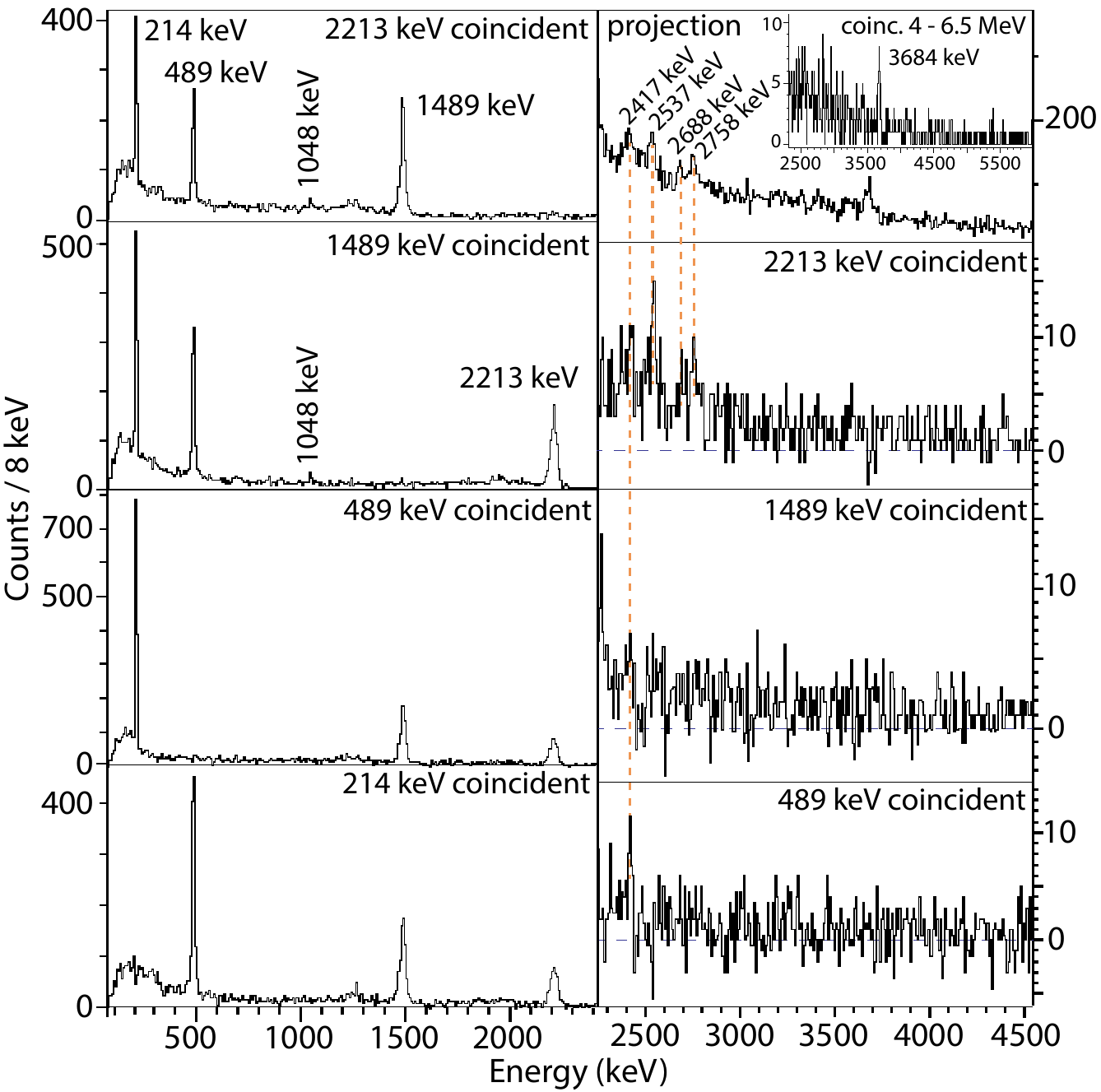}
\caption{Left: Coincidence relationships in the lower-energy part of the spectrum, with a clear level scheme emerging from these (see \ref{fig:Ca38_lev} and \cite{gad22}). Right: Doppler-corrected projection of the $\gamma\gamma$ matrix in the higher-energy region of the spectrum and coincidence spectra in this energy range obtained from cuts on the labeled transitions. Background was subtracted via a cut of equal width at slightly higher energy. Evidence for coincidence relationships is present also in the higher-energy region of the spectra. The inset shows that there may be unresolved weak high-energy transitions that feed the 3684-keV state.}
\label{fig:Ca38_coinc2}
\end{center}
\end{figure}

To construct the level scheme, $\gamma\gamma$ coincidences were used. The level scheme, 
comprising the transitions with 3\% or more relative intensity in the above Table, is 
discussed in detail in Ref. \cite{gad22}. For placement of the weaker transitions not previously discussed,  Fig. \ref{fig:Ca38_coinc2} 
shows the coincidence analysis of the higher-energy part of the \nuc{38}{Ca} spectrum (right) in addition to the clear coincidences present below 2.5~MeV (left). On the right, the figure shows the high-energy part of the projection of the $\gamma\gamma$ coincidence matrix and cuts on the three strongest $\gamma$-ray transitions in this energy region. There is evidence
that the four $\gamma$ rays between 2417 and 2758~keV are in coincidence with the 2213~keV and
that the 2417-keV transition is coincident to the 489 keV. We note that there was no evidence for
any of the high-energy $\gamma$ rays in the coincidence spectrum with 214~keV which is not
shown here. The narrow feature at 2268~keV in the spectrum coincident to 1489~keV is not visible in the singles spectrum and is not claimed as a transition here. Curiously, the 3684-keV transition, present in the projection of the coincidence
matrix, appears coincident with $\gamma$ rays between 4 and 6.5~MeV, as shown in the inset.

\begin{figure}[h]
\begin{center}
\pic[0.60]{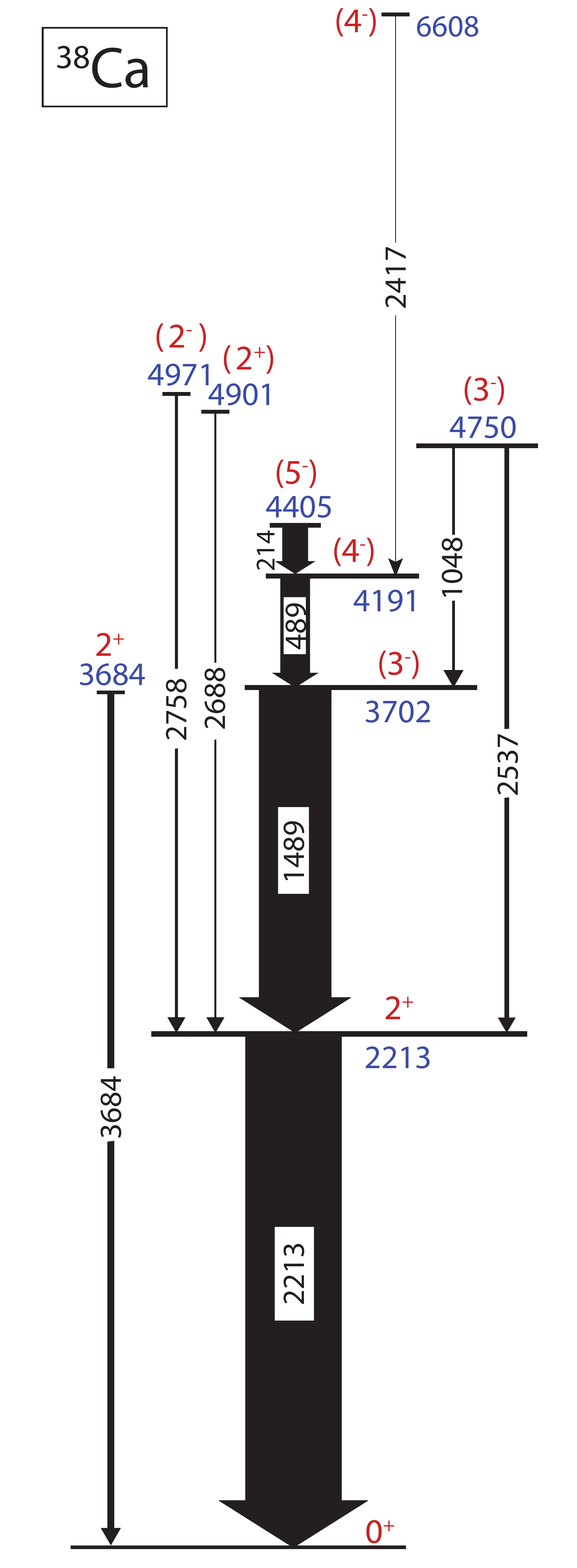}
\caption{Proposed level scheme for \nuc{38}{Ca}, extended version based on Ref. \cite{gad22}. The width of the arrows is proportional to the $\gamma$-ray intensity of the corresponding transition. The proton separation energy of $S_p=
4.54727(22)$~MeV \cite{ame20} places the four highest-excited states in this level scheme above
the proton separation energy. Spin-parity assignments in parenthesis are tentative.}
\label{fig:Ca38_lev}
\end{center}
\end{figure}

Figure \ref{fig:Ca38_lev} shows the resulting level scheme with the intensities of the
$\gamma$-ray transitions indicated by the line width of the arrows.

The 2688 and 2758-keV transitions, which appear
to feed the $2^+_1$ state, establish levels at 4901 and 4972~keV, respectively. The first
of these two levels likely corresponds to the previously reported $2^+$ at 4902~keV \cite{nndc38},
while the 4971-keV state might be the mirror level of the \nuc{38}{Ar} $(2^-$) state at
5084~keV \cite{nndc38} which decays predominantly to the first $2^+$ state, as observed here. The 2417-keV
transition on top of the ($4^-$) state suggests a level at 6608~keV in \nuc{38}{Ca}. The
only level in the mirror nucleus in that energy region, that decays the strongest to the
$4^-$ state, is another $4^-$ level at 6602~keV \cite{nndc38}. It is, however, peculiar that
the mirror level in \nuc{38}{Ar} would now be about equal in energy, while they are significantly
higher for the other negative-parity states \cite{nndc38}.

\begin{figure}[h]
\begin{center}
\pic{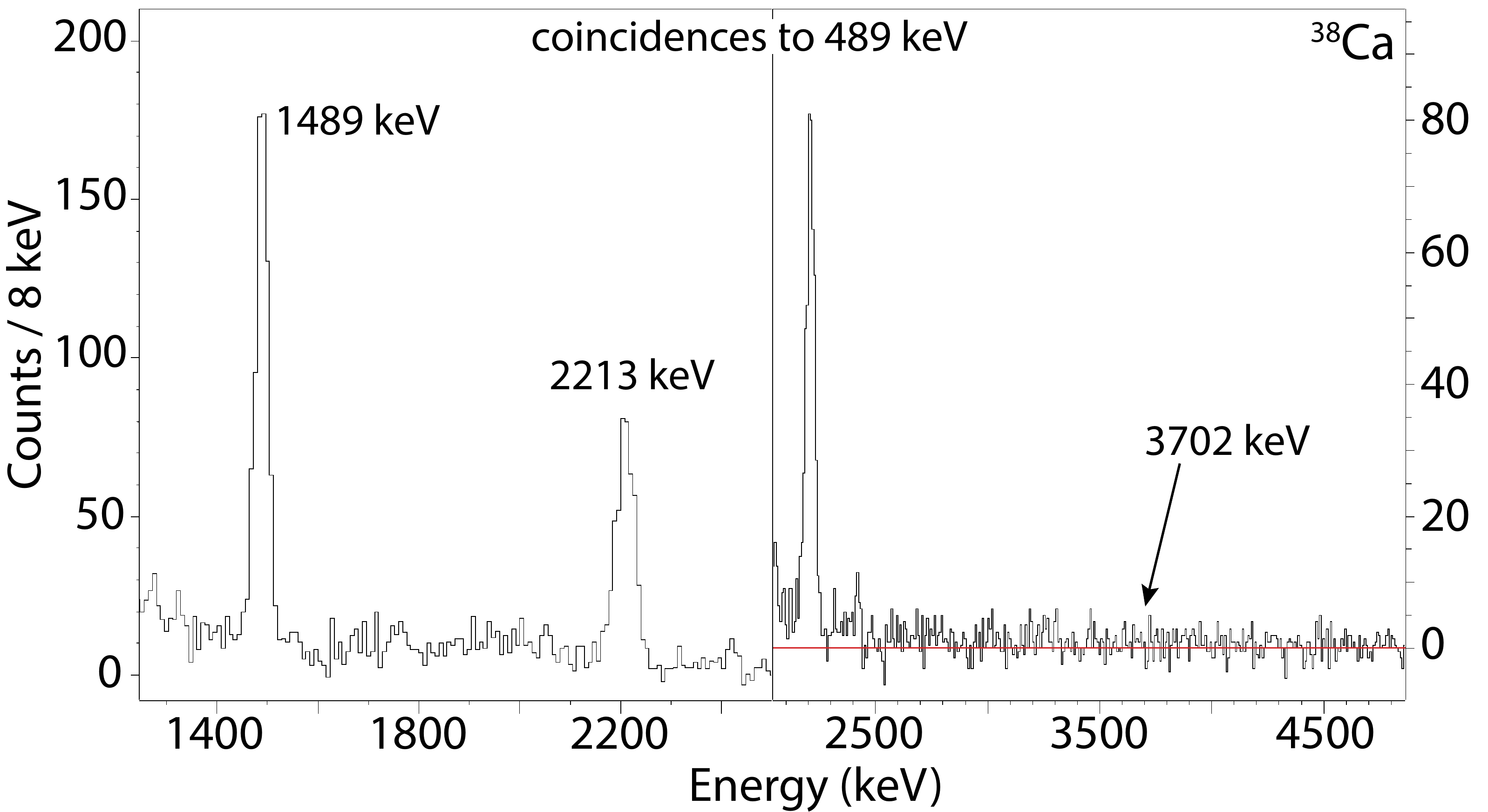}
\caption{Coincidence spectrum gated on the 489-keV feeder to the ($3^-_1$) level at 3702~keV. No
evidence for the $(3^-_1) \rightarrow 0^+_1$ transition at 3702 keV is observed.}
\label{fig:Ca38_branch}
\end{center}
\end{figure}

The present data can also shed further light on the Coulomb excitation work of Ref. \cite{cot99}
where two excited states were proposed to be excited (see Fig. 1 of \cite{cot99}), with
two transitions depopulating the 3685-keV $2^+_2$ level. The picture emerging here is rather
that, in fact, three states may be excited: the first $2^+$, the second $2^+$, and the first
$3^-$ states. The 1479-, 2206-, and 3685-keV  transitions reported in \cite{cot99} would then
correspond to the $(3^-_1) \rightarrow 2^+_1$, $2^+_1 \rightarrow 0^+_1$, and $2^+_2 \rightarrow
0^+_1$ decays, respectively. While we cannot exclude the existence of a weak $2^+_2 \rightarrow
2^+_1$ transition, due to the proximity to the spectrum-dominating 1489-keV $\gamma$ ray, as
discussed above, we can limit the $(3^-_1) \rightarrow 0^+_1$ transition that was briefly
discussed in \cite{cot99}. Figure~\ref{fig:Ca38_branch} shows a coincidence spectrum obtained
by gating on the 489-keV transition that feeds predominantly the ($3^-_1$) state.
Given the photopeak efficiencies and assuming that there are at most 3 counts at 3702~keV, we can
limit the branching ratio for the $(3^-)$ decay to the ground state to $\leq 1$\%.

Similar to the case of \nuc{39}{Ca}, discussed above, the configurations of the states 
populated in the dissipative scattering discussed here appear to require the simultaneous 
rearrangement of several nucleons in a single collision, giving hitherto unconsidered access to complex-structure 
states in rare isotopes \cite{gad22}.

\section{Summary}

In this work, the population of higher-spin complex-structure states at high momentum loss, 
discussed in Ref. \cite{gad22} for \nuc{38}{Ca} excited following inelastic scattering off 
\nuc{9}{Be}, is extended to the case of fast-beam-induced single-neutron pickup onto 
excited-state configurations of the projectile, \nuc{9}{Be}($\nuc{38}{Ca}^*, \nuc{39}{Ca}+\gamma$)X.  
The states populated in the low-momentum tail of the \nuc{39}{Ca} one-neutron pickup parallel momentum
distribution, with total angular momenta up to $J=15/2$, are the same as observed in high-spin 
spectroscopy following a fusion-evaporation reaction. There is no evidence for population of 
the low-spin states usually reported in the \nuc{40}{Ca}-induced light-ion nucleon transfer 
reactions. As proposed in Ref. \cite{gad22} for \nuc{38}{Ca}, the excited states of \nuc{39}{Ca} 
formed in one-neutron pickup onto \nuc{38}{Ca} at high momentum loss are consistent with 
the simultaneous rearrangement of multiple nucleons in a single energetic collision with a 
\nuc{9}{Be} target nucleus. It is argued that the same mechanism is responsible for the weak 
population of complex-structure states often reported in the low-momentum tails of one-nucleon 
knockout longitudinal momentum distributions. Complementing the discussion of \nuc{38}{Ca} 
in Ref. \cite{gad22}, the complete spectroscopy data on this nucleus from the present work is 
provided. While the detailed mechanism(s) responsible for the large-momentum-loss events observed 
in the reactions remain unexplained at present, these pose an interesting experimental challenge. 
Their clarification would certainly require a characterization of the target breakup channels. 
Regardless of the detailed reaction mechanism(s), this work demonstrates that exploiting such 
dissipative processes offers the potential to use high-luminosity, fast-beam-induced reactions 
to perform higher-spin spectroscopy of rare isotopes.

\begin{acknowledgments}
This work was supported by the U.S. National Science Foundation (NSF) under Grant No. PHY-1565546 and PHY-2110365,
by the DOE National Nuclear Security Administration through the Nuclear Science and Security Consortium,
under Award No. DE-NA0003180, and by the U.S. Department of Energy, Office of Science, Office of Nuclear
Physics, under Grants No. DE-SC0020451 (MSU) and DE-FG02-87ER-40316 (WashU) and under Contract No. DE-AC02-06CH11357 (ANL).  GRETINA was funded by the DOE, Office of Science. Operation of the array at NSCL was supported by the DOE under Grant No. DE-SC0019034. J.A.T. acknowledges support
from the Science and Technology Facilities Council (U.K.) Grant No. ST/V001108/1.
\end{acknowledgments}


\begin{thebibliography}{99}
\bibliographystyle{apsrev4-2}

\bibitem{dra97} G. D. Dracoulis, A. P. Byrne, T. Kibedi, T. R. McGoram and S. M. Mullins, J. Phys. G: Nucl. Part. Phys. 23 1191 (1997).

\bibitem{hae82} D.\ R.\ Haenni, T.\ T.\ Sugihara, R.\ P.\ Schmitt, G.\ Mouchaty, and U.\ Garg, Phys.\ Rev.\ C 25, 1699 (1982).

\bibitem{ide05} E. Ideguchi, M. Niikura, C. Ishida, T. Fukuchi, H. Baba, N. Hokoiwa, H. Iwasaki, T. Koike, T. Komatsubara, T. Kubo, M. Kurokawa, S. Michimasa, K. Miyakawa, K. Morimoto, T. Ohnishi, S. Ota, A. Ozawa, S. Shimoura, T. Suda, M. Tamaki, I. Tanihata, Y. Wakabayashi, K. Yoshida, and B. Cederwall, Eur. Phys. J. A 25, s01, 429 (2005).


\bibitem{bot15} S. Bottoni, S. Leoni, B. Fornal, R. Raabe, K. Rusek, G. Benzoni {\it et al.}, Phys. Rev. C 92, 024322 (2015).

\bibitem{ash21} J. Ash, H. Iwasaki, T. Mijatovic, T. Budner, R. Elder, B. Elman, M. Friedman, A. Gade, M. Grinder, J. Henderson, B. Longfellow, A. Revel, D. Rhodes, M. Spieker, Y. Utsuno, D. Weisshaar, and C. Y. Wu, Phys. Rev. C 103, L051302 (2021).

\bibitem{gad22} A. Gade, B. A. Brown, D. Weisshaar, D. Bazin, K. W. Brown, R. J. Charity,  P.
Farris,  A. M. Hill, J. Li, B. Longfellow, D. Rhodes, W. Reviol, and J. A. Tostevin, Phys. Rev. Lett. (submitted).

\bibitem{bol06} G. Bollen, D. Davies, M. Facina, J. Huikari, E. Kwan, P. A. Lofy, D. J. Morrissey, A. Prinke, R. Ringle, J. Savory, P. Schury, S. Schwarz, C. Sumithrarachchi, T. Sun, L. Weissman, Phys. Rev. Lett. 96, 152501 (2006).

\bibitem{rin07} R. Ringle, T. Sun, G. Bollen, D. Davies, M. Facina, J. Huikari, E. Kwan, D. J. Morrissey, A. Prinke, J. Savory, P. Schury, S. Schwarz, C. S. Sumithrarachchi, Phys. Rev. C 75, 055503 (2007).

\bibitem{geo07}S. George, S. Baruah, B. Blank, K. Blaum, M. Breitenfeldt, U. Hager, F. Herfurth, A. Herlert, A. Kellerbauer, H.-J. Kluge, M. Kretzschmar, D. Lunney, R. Savreux, S. Schwarz, L. Schweikhard, C. Yazidjian, Phys. Rev. Lett. 98, 162501 (2007).


\bibitem{par11} H. I. Park, J. C. Hardy, V. E. Iacob, A. Banu, L. Chen, V. V. Golovko, J. Goodwin, V. Horvat, N. Nica, E. Simmons, L. Trache, R. E. Tribble, Phys. Rev. C 84, 065502 (2011).

\bibitem{par14} H. I. Park, J. C. Hardy, V. E. Iacob, M. Bencomo, L. Chen, V. Horvat, N. Nica, B. T. Roeder, E. Simmons, R. E. Tribble, I. S. Towner, Phys. Rev. Lett. 112, 102502 (2014).

\bibitem{par15} H. I. Park, J. C. Hardy, V. E. Iacob, M. Bencomo, L. Chen, V. Horvat, N. Nica, B.T. Roeder, E. McCleskey, R. E. Tribble, I. S. Towner, Phys. Rev. C 92, 015502 (2015).

\bibitem{bla15} B. Blank, J.-C. Thomas, P. Ascher, L. Audirac, A. Bacquias, L. Caceres, G. Canchel, L. Daudin, F. de Oliveira Santos, F. Didierjean, M. Gerbaux, J. Giovinazzo, S. Grevy, T. Kurtukian Nieto, I. Matea, F. Munoz, M. Roche, L. Serani, N. Smirnova, J. Souin, Eur. Phys. J. A 51, 8 (2015).


\bibitem{lon17} A. M. Long, T. Adachi, M. Beard, G. P. A. Berg, Z. Buthelezi, J. Carter, M. Couder, R. J. deBoer, R. W. Fearick, S. V. Fortsch, J. Gorres, J. P. Mira, S. H. T. Murray, R. Neveling, P. Papka, F. D. Smit, E. Sideras-Haddad, J. A. Swartz, R. Talwar, I. T. Usman, M. Wiescher, J. J. Van Zyl, A. Volya, Phys. Rev. C 95, 055803 (2017).

\bibitem{boh77} W. Bohne, K. D. Buchs, H. Fuchs, K. Grabisch, D. Hilscher, U. Jahnke, H. Kluge, T. G. Masterson, H. Morgenstern, Nucl. Phys. A284, 14 (1977).

\bibitem{alf86} W. P. Alford, P. Craig, D. A. Lind, R. S. Raymond, J. Ullman, C. D. Zafiratos, B. H. Wildenthal, Nucl. Phys. A457, 317 (1986).

\bibitem{har66} J. C. Hardy, D. J. Skyrme, I. S. Towner, Phys. Lett. 23, 487 (1966).

\bibitem{kub74} S. Kubono, S. Kato, M. Yasue, H. Ohnuma, K. Ogawa, Phys. Lett. 49B, 37 (1974).

\bibitem{kub77} S. Kubono, S. Kato, M. Yasue, H. Ohnuma, M. Sasao, K. Tsukamoto, R. Kuramasu, Nucl. Phys. A276, 201 (1977).

\bibitem{pad72} R. A. Paddock, Phys. Rev. C5, 485 (1972).


\bibitem{cot99} P. D. Cottle, M. Fauerbach, T. Glasmacher, R. W. Ibbotson, K. W. Kemper, B. Pritychenko, H. Scheit, M. Steiner, Phys. Rev. C 60, 031301(R) (1999).

\bibitem{sha70} M. H. Shapiro, A. Adams, C.Moss, W. M. Denny, Nucl. Phys. A144, 17 (1970).


\bibitem{nndc38} J. Chen, Nuclear Data Sheets 152, 1 (2018).

\bibitem{hal20} M. R. Hall, D. W. Bardayan, T. Baugher, A. Lepailleur, S. D. Pain, A. Ratkiewicz, S. Ahn, J. M. Allen, J. T. Anderson, A. D. Ayangeakaa, J. C. Blackmon, S. Burcher, M. P. Carpenter, S. M. Cha, K. Y. Chae, K. A. Chipps, J. A. Cizewski, M. Febbraro, O. Hall, J. Hu, C. L. Jiang, K. L. Jones, E. J. Lee, P. D. O'Malley, S. Ota, B. C. Rasco, D. Santiago-Gonzalez, D. Seweryniak, H. Sims, K. Smith, W.P. Tan, P. Thompson, C. Thornsberry, R. L. Varner, D. Walter, G. L. Wilson, S. Zhu, Phys. Rev. C 101, 015804 (2020).

\bibitem{and99} Th. Andersson, D. Rudolph, C. Baktash, J. Eberth, C. Fahlander, D. Haslip, D. R. LaFosse, S. D. Paul, D. G. Sarantites, C. E. Svensson, H. G. Thomas, J. C. Waddington, W. Weintraub, J. N. Wilson, B. A. Brown, Eur. Phys. J. A 6, 5 (1999).


\bibitem{gad20} A. Gade, D. Weisshaar, B. A. Brown, J. A. Tostevin, D. Bazin, K. Brown, R. J. Charity, P. J. Farris, A. M. Hill, J. Li, B. Longfellow, W. Reviol, D. Rhodes, Phys. Lett. B 808, 135637 (2020).

\bibitem{nscl} A. Gade and B. M. Sherrill, Phys. Scr. 91, 053003 (2016).

\bibitem{a1900} D.\ J.\ Morrissey, B.\ M.\ Sherrill, M. Steiner, A. Stolz, and I. Wiedenh\"over, Nucl.\ Instrum.\ Methods
  in Phys.\ Res.\ B 204, 90 (2003).

\bibitem{pas13} S. Paschalis et al., Nucl. Instrum. Methods Phys. Res., Sect. A 709, 44 (2013).

\bibitem{wei17} D. Weisshaar, D. Bazin, P. C. Bender, C. M. Campbell, F. Recchia, V. Bader, T. Baugher, J. Belarge, M. P. Carpenter, H. L. Crawford, M. Cromaz, B. Elman, P. Fallon, A. Forney, A. Gade, J. Harker, N. Kobayashi, C. Langer, T. Lauritsen, I. Y. Lee, A. Lemasson, B. Longfellow, E. Lunderberg, A. O. Macchiavelli, K. Miki, S. Momiyama, S. Noji, D. C. Radford, M. Scott, J. Sethi, S. R. Stroberg, C. Sullivan, R. Titus, A. Wiens, S. Williams, K. Wimmer, and S. Zhu, Nuclear Instrum. Methods Phys. Res. Sect. A 847, 187 (2017).

\bibitem{s800} D.\ Bazin, J.\ A.\ Caggiano, B.\ M.\  Sherrill, J. Yurkon, and A. Zeller, Nucl.\ Instrum.\ Methods in Phys.\
  Res.\ B 204, 629 (2003).

\bibitem{s800FP} J. Yurkon, D. Bazin, W. Benenson, D. J. Morrissey, B. M. Sherrill, D. Swan, R. Swanson, Nucl.\ Instrum.\ Methods in Phys.\  Res.\ A 422, 291 (1999).

\bibitem{gad11} A. Gade, J. A. Tostevin, T. Baugher, D. Bazin, B. A. Brown, C. M. Campbell, T. Glasmacher, G. F. Grinyer, S. McDaniel, K. Meierbachtol, A. Ratkiewicz, S. R. Stroberg, K. A. Walsh, D. Weisshaar, and R. Winkler, Phys. Rev. C 83, 054324 (2011).

\bibitem{gad07} A. Gade, P. Adrich, D. Bazin, M. D. Bowen, B. A. Brown, C. M. Campbell, J. M. Cook, T. Glasmacher, K. Hosier, S. McDaniel, D. McGlinchery, A. Obertelli, L. A. Riley, K. Siwek, J. A. Tostevin, and D. Weisshaar, Phys. Rev. C 76 , 061302(R) (2007).

\bibitem{gad08} A. Gade, P. Adrich, D. Bazin, M.D. Bowen, B. A. Brown, C. M. Campbell, J. M. Cook, T. Glasmacher, K. Hosier, S. McDaniel, D. McGlinchery, A. Obertelli, L. A. Riley, K. Siwek, J. A. Tostevin, and D. Weisshaar, Phys. Lett. B 666, 218 (2008).

\bibitem{mcd12} S. McDaniel, A. Gade, J. A. Tostevin, T. Baugher, D. Bazin, B. A. Brown, J. M. Cook, T. Glasmacher, G. F. Grinyer, A. Ratkiewicz, and D. Weisshaar, Phys. Rev. C 85, 014610 (2012).

\bibitem{gad16} A. Gade, J. A. Tostevin, V. Bader, T. Baugher, D. Bazin, J. S. Berryman, B. A. Brown, D. J. Hartley, E. Lunderberg, F. Recchia, S. R. Stroberg, Y. Utsuno, D. Weisshaar, and K. Wimmer, Phys. Rev. C 93, 031601(R) (2016).

\bibitem{sou92} G.\ A.\ Souliotis, D.\ J.\ Morrissey, N.\ A.\ Orr, B.\ M.\ Sherrill, and J.\ A.\ Winger, Phys.\ Rev.\ C 46, 1383 (1992).


\bibitem{str14} S. R. Stroberg, A. Gade, J. A. Tostevin, V. M. Bader, T. Baugher, D. Bazin, J. S. Berryman, B. A. Brown, C. M. Campbell, K. W. Kemper, C. Langer, E. Lunderberg, A. Lemasson, S. Noji, F. Recchia, C. Walz, D. Weisshaar, and S. J. Williams, Phys. Rev. C 90, 034301 (2014).

\bibitem{mut16} A. Mutschler, O. Sorlin, A. Lemasson, D. Bazin, C. Borcea, R. Borcea, A. Gade, H. Iwasaki, E. Khan, A. Lepailleur, F. Recchia, T. Roger, F. Rotaru, M. Stanoiu, S. R. Stroberg, J. A. Tostevin, M. Vandebrouck, D. Weisshaar, and K. Wimmer, Phys. Rev. C 93, 034333 (2016).

\bibitem{cra17} H. L. Crawford, A. O. Macchiavelli, P. Fallon, M. Albers, V. M. Bader, D. Bazin, C. M. Campbell, R. M. Clark, M. Cromaz, J. Dilling, A. Gade, A. Gallant, J. D. Holt, R. V. F. Janssens, R. Kr\"ucken, C. Langer, T. Lauritsen, I. Y. Lee, J. Menendez, S. Noji, S. Paschalis, F. Recchia, J. Rissanen, A. Schwenk, M. Scott, J. Simonis, S. R. Stroberg, J. A. Tostevin, C. Walz, D. Weisshaar, A. Wiens, K. Wimmer, and S. Zhu, Phys. Rev. C 95, 064317 (2017).

\bibitem{spi19} M. Spieker, A. Gade, D. Weisshaar, B. A. Brown, J. A. Tostevin, B. Longfellow, P. Adrich, D. Bazin, M. A. Bentley, J. R. Brown, C. M. Campbell, C. Aa. Diget, B. Elman, T. Glasmacher, M. Hill, B. Pritychenko, A. Ratkiewicz, and D. Rhodes, Phys. Rev. C 99, 051304(R) (2019).


\bibitem{nndc39} J. Chen, Nuclear Data Sheets 149, 1 (2018).

\bibitem{gad19} A. Gade, R. V. F. Janssens, J. A. Tostevin, D. Bazin, J. Belarge, P. C. Bender, S. Bottoni, M. P. Carpenter, B. Elman, S. J. Freeman, T. Lauritsen, S. M. Lenzi, B. Longfellow, E. Lunderberg, A. Poves, L. A. Riley, D. K. Sharp, D. Weisshaar, and S. Zhu, Phys. Rev. C 99, 011301(R) (2019).

\bibitem{ame20} M. Wang, W. J. Huang, F. G. Kondev, G. Audi, and S. Naimi, Chin. Phys. C 45, 030003 (2021).

\bibitem{set18} K. Setoodehnia, C. Marshall, J. H. Kelley, J. Liang, F. Portillo Chaves, and R. Longland, Phys. Rev. C 98, 055804 (2018).

\bibitem{lub19} R. S. Lubna, K. Kravvaris, S. L. Tabor, Vandana Tripathi, A. Volya, E. Rubino, J. M. Allmond, B. Abromeit, L. T. Baby, and T. C. Hensley, Phys. Rev. C 100, 034308 (2019).

\bibitem{lub20} R. S. Lubna, K. Kravvaris, S. L. Tabor, Vandana Tripathi, E. Rubino, and A. Volya, Phys. Rev. Research 2, 043342 (2020).



\end{thebibliography}
\end{document}